\documentclass[reprint,aps,epsfig,superscriptaddress,twocolumn]{revtex4-1}
\usepackage{bm}
\usepackage[dvips]{graphicx}
\usepackage{mathrsfs}
\usepackage[intlimits]{amsmath}
\usepackage{textcomp}
\usepackage[switch]{lineno}
\usepackage[colorlinks, citecolor=red]{hyperref}
\usepackage{tabularx}

\begin{document}

\title{Heralded ion-atom entanglement in a hybrid quantum network}

\author{P.-Y. Liu}
\altaffiliation{These authors contributed equally to this work}
\affiliation{Center for Quantum Information, Institute for Interdisciplinary Information Sciences, Tsinghua University, Beijing 100084, PR China}

\author{Y.-B. Liang}
\altaffiliation{These authors contributed equally to this work}
\affiliation{Center for Quantum Information, Institute for Interdisciplinary Information Sciences, Tsinghua University, Beijing 100084, PR China}

\author{Z.-B. Cui}
\altaffiliation{These authors contributed equally to this work}
\affiliation{Center for Quantum Information, Institute for Interdisciplinary Information Sciences, Tsinghua University, Beijing 100084, PR China}

\author{J.-X. Shi}
\altaffiliation{These authors contributed equally to this work}
\affiliation{Center for Quantum Information, Institute for Interdisciplinary Information Sciences, Tsinghua University, Beijing 100084, PR China}

\author{Y.-D. Sun}
\affiliation{Center for Quantum Information, Institute for Interdisciplinary Information Sciences, Tsinghua University, Beijing 100084, PR China}

\author{Z.-C. Tian}
\affiliation{Center for Quantum Information, Institute for Interdisciplinary Information Sciences, Tsinghua University, Beijing 100084, PR China}

\author{H.-S. Sun}
\affiliation{Center for Quantum Information, Institute for Interdisciplinary Information Sciences, Tsinghua University, Beijing 100084, PR China}

\author{Y. Wang}
\affiliation{HYQ Co., Ltd., Beijing 100176, PR China}

\author{L.-M. Duan}
\email{lmduan@tsinghua.edu.cn}
\affiliation{Center for Quantum Information, Institute for Interdisciplinary Information Sciences, Tsinghua University, Beijing 100084, PR China}
\affiliation{Hefei National Laboratory, Hefei 230088, PR China}

\author{Y.-F. Pu}
\email{puyf@tsinghua.edu.cn}
\affiliation{Center for Quantum Information, Institute for Interdisciplinary Information Sciences, Tsinghua University, Beijing 100084, PR China}
\affiliation{Hefei National Laboratory, Hefei 230088, PR China}

\begin{abstract}
Quantum networks, which interconnect local nodes via photonic channels, represent a promising approach to scale up quantum communication and computation systems. However, their development has been largely restricted to homogeneous architectures employing a single type of matter qubit, precluding the exploitation of the complementary strengths inherent to distinct physical platforms. In this work, we experimentally realize a prototype hybrid quantum network by linking a trapped-ion node with an atomic-ensemble node separated by a distance of $4\,$m through photonic channels. By employing techniques such as photonic envelope shaping and dual-color Bell-state measurements, we establish heralded entanglement between these two remote heterogeneous nodes, which makes a foundational step for hybrid quantum networks and distributed quantum information systems. The resulting entanglement fidelity significantly surpasses the classical bound, validating the effectiveness of the hybrid interconnection. Our work provides a key building block for future large-scale hybrid quantum networks.
\end{abstract}

\maketitle

\subsection*{Introduction}

Quantum networks based on repeaters~\cite{kimble_internet,BDCZ,DLCZ,duan_arxiv} represent a powerful approach to scaling up the size, extending the spatial reach, and enhancing the performance of current quantum information systems. This distributed architecture enables important applications such as distributed quantum computing~\cite{distributed, duan_arxiv, duan_pra,oxford_distributed, rempe, duan_RMP}, networked quantum sensing~\cite{ye_and_lukin,repeater_telescope, oxford_clock, lukin_nv_telescope}, and long-distance quantum communication~\cite{BDCZ,DLCZ,rmp_gisin,oxford_qkd,weifurner_qkd, pan_ion, pan_atom}. Quantum networks and repeaters have been demonstrated in various physical platforms~\cite{pan_ion, pan_atom,lukin_nv_telescope,oxford_distributed,rempe,weifurner_qkd,2021nature,230m,hanson_long, 500Hz, multiplexed,bao_remote,sc_entanglement, bao_420km}.

The performance of a quantum network relies heavily on the capabilities of its constituent matter qubits---specifically, their ability to store and process local quantum information and to interface with flying qubits, which varies significantly across different physical platforms. For instance, single-spin systems typically exhibit superior performance in quantum logic gates, whereas spin-ensemble systems generally feature stronger interactions with single photons. However, conventional quantum networks have predominantly relied on a single type of matter qubit, thereby lacking the ability to harness the complementary strengths of different physical platforms. Furthermore, the envisioned future global quantum internet~\cite{ACM} must ultimately possess the capability to interconnect diverse quantum devices---typically based on different physical platforms---analogous to the classical global internet. A hybrid architecture capable of harnessing distinct physical systems therefore emerges as a vital solution to these challenges~\cite{hybrid_proposal}.

In building a hybrid quantum network, the foundational step is distributing heralded entanglement between two remote heterogeneous nodes. While some recent works have demonstrated quantum state transfer or post-selected entanglement between distinct matter systems by emitting a photon from one node and absorbing it at another~\cite{riedmatten_hybrid,zou_hybrid,bernien_hybrid}, these implementations lack a crucial heralding signal. Consequently, the resulting entangled states are dominated by vacuum components, limiting their utility and scaling for subsequent tasks such as quantum teleportation~\cite{rempe_teleportation,hanson_teleportation}, entanglement purification~\cite{hanson_purification}, and distributed computing~\cite{duan_arxiv,distributed,oxford_distributed}.

In this work, we experimentally realize a prototype hybrid quantum network by establishing heralded entanglement between two distinct but complementary physical platforms: a trapped ion and an atomic ensemble. The former offers superior logic gate fidelities~\cite{oxford_gate, quantiuum} and long memory coherence~\cite{xuyulin,kihwan}, while the latter currently holds the record for entanglement distribution distance and rate~\cite{bao_remote, bao_420km, 500Hz}. A primary challenge in entangling these two platforms lies in the mismatched frequencies and envelope shapes of the photons emitted from their respective physical systems. In this experiment, we precisely shape the wavepacket of the photon emitted from the trapped-ion node to match the envelope of the photon from the atomic-ensemble node. Additionally, we upconvert the $780\,$nm photon from the atomic-ensemble node to $397\,$nm via sum-frequency generation (SFG), thereby matching the optical wavelength of the trapped-ion node. Finally, through the development of a dual-color Bell state measurement (BSM), we successfully establish heralded ion-atom entanglement mediated by photon interference. We believe that the heralded ion-atom entanglement and the technologies developed in this work will serve as a vital building block for future hybrid quantum networks.

\begin{figure*}[]
    \centering
    \includegraphics[width=1.0\textwidth]{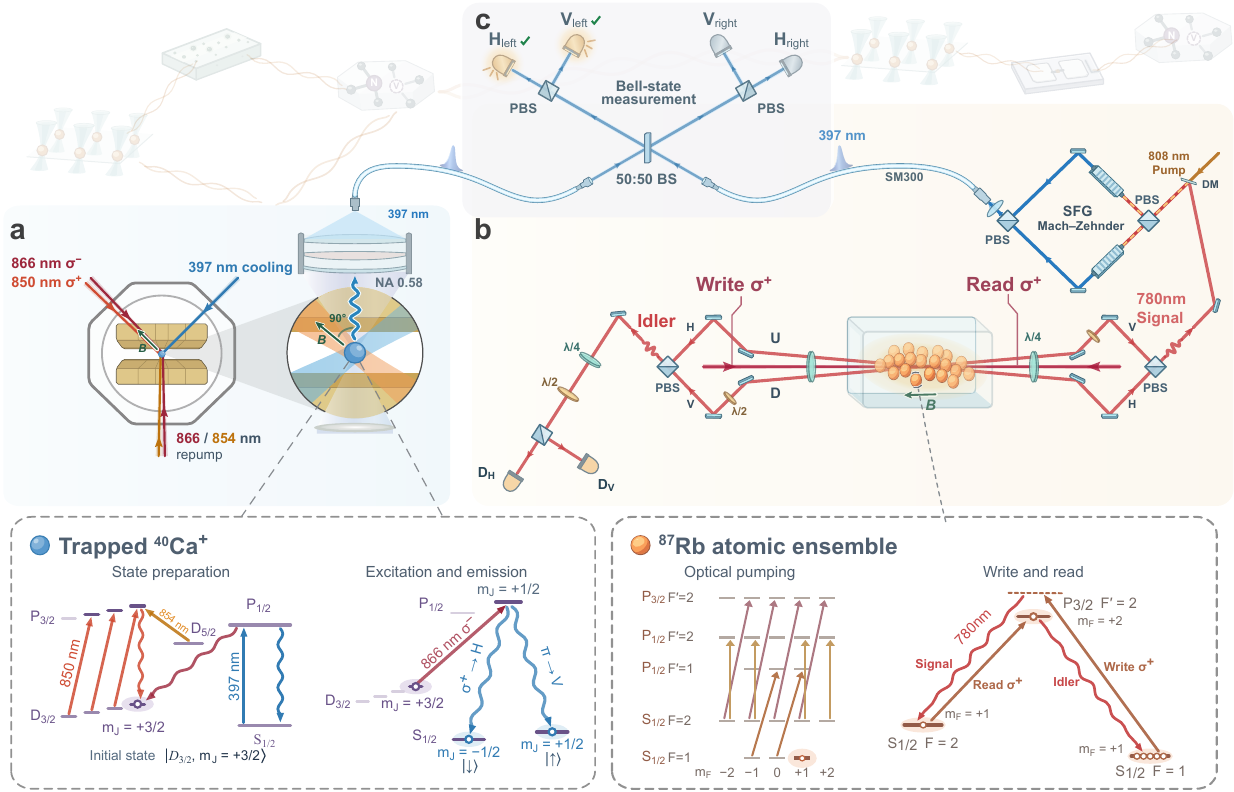}
    \caption{
    \textbf{Ion-atom hybrid quantum network.}
    A prototype ion-atom quantum network is enabled by distributing heralded entanglement between a trapped-ion node and an atomic-ensemble node. The upper schematic illustrates an envisioned hybrid quantum network interconnecting diverse quantum devices across distinct physical platforms.
    \textbf{a}, Trapped-ion node. A single $^{40}\mathrm{Ca}^+$ ion in a segmented blade trap is first initialized into the $|D_{3/2}, m_J = +3/2\rangle$ state via optical pumping and then excited to $|P_{1/2}, m_J = +1/2\rangle$ by an envelope-shaped $\sigma^-$-polarized 866\,nm pulse (lower-left inset). By collecting the spontaneously emitted 397\,nm photon perpendicular to the magnetic field $B$, a maximally entangled ion-photon state is generated.
    \textbf{b}, Atomic-ensemble node and polarization-preserving wavelength conversion. Using the Duan--Lukin--Cirac--Zoller scheme in optically pumped $^{87}\mathrm{Rb}$ atoms (lower-right inset), a weak write pulse generates two forward-scattered 780\,nm signal modes that are combined at a PBS to yield atom-photon entanglement. The 780\,nm photon is upconverted to 397\,nm via SFG with an 808\,nm pump laser in an actively phase-locked Mach--Zehnder interferometer incorporating a PPLN waveguide in each arm. A read pulse retrieves the atomic memory modes into an idler photon, which is later measured via waveplates, a PBS and detectors $D_H$ and $D_V$.
    \textbf{c}, Dual-color Bell-state measurement. Envelope-matched 397\,nm photons from both nodes undergo photon interference at a 50:50 BS. Coincidence detection of $\mathrm{H}_{\mathrm{left}}$ and $\mathrm{V}_{\mathrm{left}}$ after the two PBSs heralds ion-atom entanglement. BS, beam splitter; DM, dichroic mirror; NA, numerical aperture; PBS, polarizing beam splitter; PPLN, periodically poled lithium niobate; SFG, sum-frequency generation; $\lambda/2$, half-wave plate; $\lambda/4$, quarter-wave plate.
    }
    \label{FIG1}
\end{figure*}

\subsection*{Spin-photon entanglement}

\begin{figure*}[]
    \centering
    \includegraphics[width=0.85\textwidth]{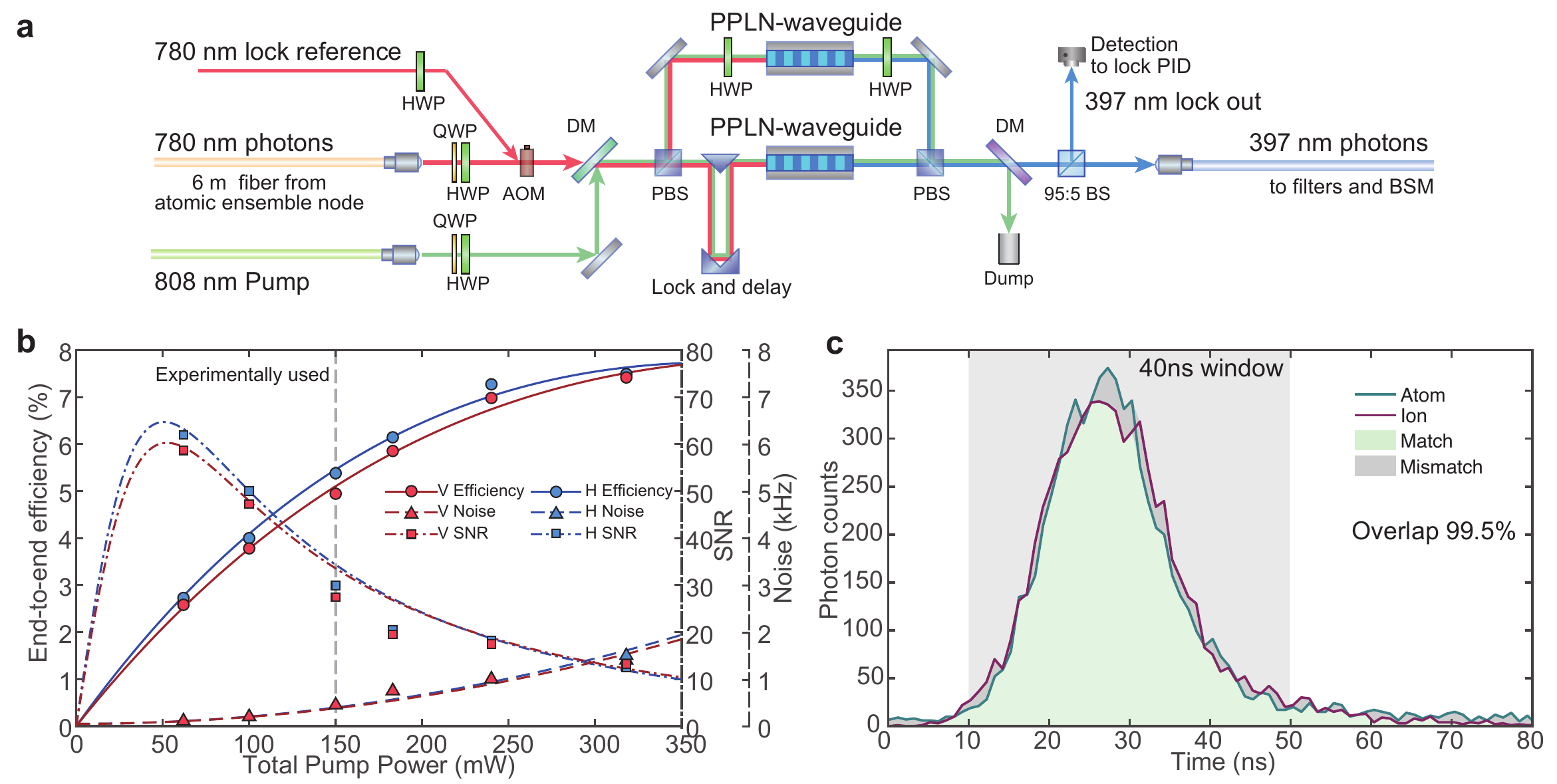}
    \caption{\textbf{Polarization-preserving frequency conversion and wavepacket shaping.}
    \textbf{a}, Schematic of the quantum frequency conversion (QFC) setup. Polarization-encoded $780\,$nm signal photons from the Rb atomic ensemble are combined with an $808\,$nm pump laser at a dichroic mirror and directed into a Mach-Zehnder (MZ) interferometer containing a pair of periodically poled lithium niobate (PPLN) waveguides. The waveguides in the two arms separately convert the horizontal ($H$) and vertical ($V$) polarization components to $397\,$nm. These converted components are recombined at an output polarizing beam splitter (PBS) and then routed to the Bell-state measurement (BSM) setup. To actively lock the relative phase between the two interferometer arms, a $780\,$nm reference laser is coupled into the single-photon path via an acousto-optic modulator (AOM) during idle periods of the experimental cycle (e.g., during laser cooling). After conversion to $397\,$nm, the reference light is sampled by a $95$:$5$ beam splitter (BS) for polarization projection measurements, and the result is fed back in real time to a piezoelectric transducer in one arm to stabilize the path phase.
    \textbf{b}, Performance of the QFC module. The end-to-end efficiency, background noise, and signal-to-noise ratio (evaluated for the atomic ensemble operating at a $1\%$ excitation probability within a $40\,$ns window) of the QFC module are characterized as a function of the total $808\,$nm pump power. The vertical dashed line labeled `Experimentally used' marks the operating point chosen for the experiment. At this pump power, the end-to-end efficiency (accounting for frequency conversion, all fiber coupling, and all filtering stages) is $5.4\%$ for the $H$ arm and $4.9\%$ for the $V$ arm, yielding SNRs of $30$ and $27$, respectively.
    \textbf{c}, Temporal profiles of single photons at 397~nm. We measure the temporal envelopes of single photons emitted by the ion and frequency-converted photons from the atomic ensemble. Both wavepackets are carefully tailored to achieve optimal indistinguishability. Using the overlap integral $\chi=\dfrac{\left(\int_{t_1}^{t_2}\sqrt{y_{\mathrm{ion}}(t)\,y_{\mathrm{atom}}(t)}\,\mathrm{d}t\right)^{2}}{\left(\int_{t_1}^{t_2}y_{\mathrm{ion}}(t)\,\mathrm{d}t\right)\left(\int_{t_1}^{t_2}y_{\mathrm{atom}}(t)\,\mathrm{d}t\right)}$, we obtain a temporal mode overlap of $99.5\%$ within the $40\,$ns acceptance time window.
    }
    \label{FIG2}
\end{figure*}

We build a trapped-ion quantum network node and an atomic-ensemble quantum network node within the same laboratory separated by a distance of $4\,$m. A single $^{40}$Ca$^+$ ion is trapped and laser-cooled in a segmented blade trap, with trap frequencies of $\{\omega_x,\omega_y,\omega_z\}=2\pi\times\{3.1,3.3,0.4\}\,$MHz~\cite{multiplexed}. The ion is initialized into the state $|D_{3/2},m=+3/2\rangle$ via a combination of a $\sigma^+$-polarized $850\,$nm beam, an $854\,$nm beam, and a $397\,$nm beam. Subsequently, a short $\sigma^{-}$-polarized $866\,$nm pulse is applied to excite ion-photon entanglement. This $866\,$nm laser pulse drives the transition from the initial state to the excited state $|P_{1/2},m=+1/2\rangle$. Following the spontaneous decay from the excited state, the emitted $397\,$nm photon is collected in a direction perpendicular to the magnetic field. Under this configuration, the resulting ion-photon entangled state can be expressed as:
\begin{equation}
|\Psi_{\text{IP}}\rangle=\frac{1}{\sqrt{2}}(|\downarrow\rangle|H\rangle_\text{I} + |\uparrow\rangle|V\rangle_\text{I})
\end{equation}
where $|\downarrow\rangle\equiv|S_{1/2},m=-1/2\rangle$/$|\uparrow\rangle\equiv|S_{1/2},m=+1/2\rangle$ denotes the corresponding Zeeman sublevel of the ground state and $|H\rangle_\text{I}$/$|V\rangle_\text{I}$ represents the horizontal/vertical polarization of the photon emitted from the trapped-ion node.

\begin{figure*}[]
    \centering
    \includegraphics[width=0.75\textwidth]{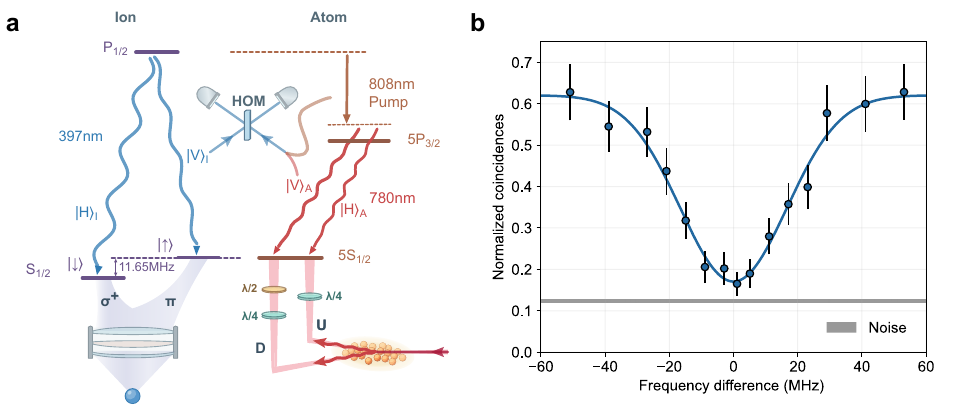}
    \caption{\textbf{Frequency matching and Hong--Ou--Mandel (HOM) interference.}
    \textbf{a}, Structure of spin-photon entanglement and frequency-matching scheme. The $^{40}\mathrm{Ca}^+$ ion decays from $|P_{1/2}, m_J = +1/2\rangle$ and emits a $397\,$nm photon in either $H$ or $V$ polarization. Here $|H\rangle_\text{I}$ and $|V\rangle_\text{I}$ differ by $11.65\,\mathrm{MHz}$ due to the Zeeman effect. In contrast, the two polarization components $|H\rangle_\text{A}$ and $|V\rangle_\text{A}$ from the $^{87}\mathrm{Rb}$ atomic ensemble have the same frequency. The dual-color Bell state measurement necessitates that the polarization components $|V\rangle_\text{I}$, $|H\rangle_\text{A}$, and $|V\rangle_\text{A}$ share a common frequency after frequency conversion. We carefully tune the frequency of the $808\,$nm pump laser in the wavelength conversion module and perform HOM interference to achieve this.
    \textbf{b,} HOM interference. Here we pick out the two vertical photon components $|V\rangle_\text{I}$ and $|V\rangle_\text{A}$ from both nodes and direct them into a HOM interferometer. The coincidence events between the two BS output ports are measured as a function of the photon frequency difference, with a $40\,$ns acceptance window. The grey band ($0.124 \pm 0.004$) indicates the theoretical noise background arising from double excitations in the atomic ensemble and coincidences between QFC background photons and those from the ion. The data exhibit a full width at half maximum (FWHM) of $39\,\mathrm{MHz}$. This result enables the frequency matching required in the dual-color Bell state measurement and characterizes the indistinguishability of photons from the two sources.
    }
    \label{FIG3}
\end{figure*}

The atomic-ensemble node is built with $^{87}$Rb atoms. Here we utilize the DLCZ (Duan-Lukin-Cirac-Zoller) scheme~\cite{DLCZ} to generate the atom-photon entanglement. We first prepare the atoms in the initial state $|5S_{1/2},F=1,m_F=+1\rangle$ via optical pumping. By exciting the atomic cloud with a weak write pulse and combining two forward-scattered signal modes, U and D, at a PBS (polarizing beam splitter)~\cite{lattice}, we generate an atom-photon entangled state described as:
\begin{equation}
|\Psi_{\text{AP}}\rangle=\frac{1}{\sqrt{2}}(|U\rangle|H\rangle_\text{A} + |D\rangle|V\rangle_\text{A})
\end{equation}
where $|U\rangle$/$|D\rangle$ denotes the collective atomic memory in spatial mode U/D and $|H\rangle_\text{A}$/$|V\rangle_\text{A}$ represents the horizontal/vertical polarization of the signal photon after the PBS. The strategies and energy levels used for state preparation and excitation in both nodes are illustrated in Fig.~1.

\subsection*{Frequency and envelope matching}

 In implementing a hybrid quantum network, a critical obstacle lies in the large frequency discrepancy ($>$THz) between the optical transitions employed by different physical platforms. To enable ion-atom entanglement through photon interference, we upconvert the $780\,$nm photon from the atomic node to $397\,$nm to match the frequency of the ionic node. Here we use frequency up-conversion rather than down-conversion~\cite{maryland} to suppress the influence of double excitation events in the atomic ensemble node. We employ a pair of PPLN (periodically poled lithium niobate) waveguides in a Mach-Zehnder interferometer for the polarization-preserving wavelength conversion~\cite{MZ_ppln}, as shown in Fig.~2a. The relative phase between the two interferometer arms is actively locked to guarantee a stable phase of the atom-photon entanglement after the conversion. We achieve an end-to-end efficiency of abou $5\%$ (device efficiency of nearly $30\%$) and a signal-to-noise ratio of $30$ with an excitation probability of about $1\%$ in the atomic node and a $40\,$ns acceptance window, as shown in Fig.~2b.

Another issue influencing the quality of the photon interference is the different temporal envelopes of the photons emitted from different systems. The $^{40}$Ca$^+$ ion and the $^{87}$Rb atom have excited state lifetimes of $6.9\,$ns and $26\,$ns, respectively~\cite{Ca,Rb}. Thus the photon from the atomic node has a longer wavepacket. Here we shape the temporal envelope of the photon from the ion by carefully adjusting the intensity, detuning and temporal shape of the $866\,$nm excitation laser~\cite{blinov_shaping,Laurat_shaping}, and achieve a high wavepacket overlap for photons from both sides, as shown in Fig.~2c. The overlap integral is measured to be $99.5\%$ for an acceptance window of $40\,$ns, confirming the good matching of the photon envelopes.

\subsection*{Dual-color Bell state measurement}

\begin{figure*}[]
    \centering
    \includegraphics[width=0.78\textwidth]{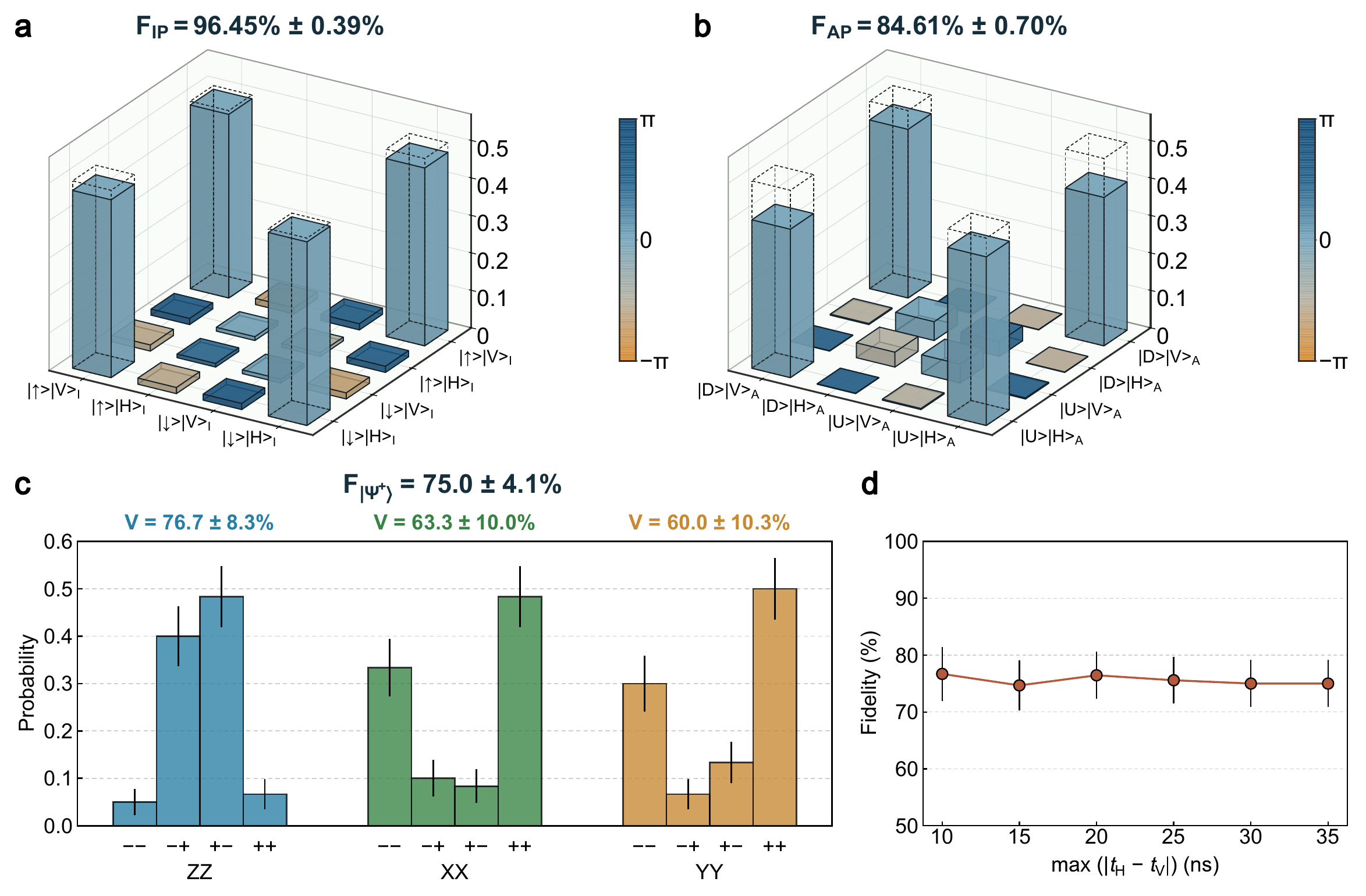}
    \caption{\textbf{Heralded ion-atom entanglement.}
    \textbf{a, b}, Reconstructed density matrices of the ion-photon entangled state of the $^{40}\mathrm{Ca}^+$ ion
    and the atom-photon entangled state of the $^{87}\mathrm{Rb}$ atomic ensemble after quantum frequency conversion of the signal photon, respectively. In both cases, photon-state measurements are performed at the central Bell-state measurement (BSM) station.
    \textbf{c}, Measurement results of correlations in three bases $XX$, $YY$ and $ZZ$ used to determine the fidelity of the heralded
    ion--atom entangled state. A total of $180$ heralded entanglement events were recorded.
    \textbf{d}, Ion-atom entanglement fidelity as a function of the maximally allowed arrival-time difference between the $H$- and $V$-polarized photons, $\max(|t_H-t_V|)$. The fidelity remains consistent within statistical uncertainties across the investigated ranges.
    }
    \label{FIG4}
\end{figure*}

A further challenge in building a hybrid quantum network is that the spin-photon entangled states from two heterogeneous nodes typically lack the symmetric structure found in a homogeneous quantum network. For the generated ion-photon entanglement (Eq.~(1)), there exists an $11.65\,$MHz frequency splitting between the two internal states $|\downarrow\rangle$ and $|\uparrow\rangle$; thus, the horizontal component of the photon $|H\rangle_\text{I}$ is $11.65\,$MHz higher in frequency than the vertical component $|V\rangle_\text{I}$. In contrast, for the atom-photon entanglement, the horizontal component $|H\rangle_\text{A}$ has the same frequency as the vertical component $|V\rangle_\text{A}$. The energy level structures and corresponding photon frequencies for both nodes are illustrated in Fig.~3. During a standard Bell state measurement, this frequency asymmetry can induce a time-dependent phase arising from the random photon detection events, thus deteriorating the fidelity of the heralded ion-atom entanglement.

We address this issue via a dual-color Bell-state measurement. We first match the frequency of the vertical components of the two photons from both sides by carefully tuning the frequency of the $808\,$nm pump laser in the wavelength conversion module. We send the vertical components of the photons from both sides onto a beamsplitter and observe the Hong-Ou-Mandel (HOM) interference~\cite{HOM}. The clear HOM dip confirms that the frequencies of the vertical components and the temporal envelopes are in good agreement for photons from both sides, as shown in Fig.~3. Subsequently, we implement the Bell state measurement by sending both polarization components of the photons to the BSM setup. Here we select one of the four Bell states, $|\Psi^+\rangle=(|H\rangle_\text{I}|V\rangle_\text{A}+|V\rangle_\text{I}|H\rangle_\text{A})/\sqrt{2}$, as a successful heralding event triggered by the coincidence detection of detectors H$_\text{left}$ and V$_\text{left}$ as shown in Fig.~1. The heralded ion-atom entanglement can be expressed as:
\begin{equation}
|\Psi_{\text{IA}}\rangle=\frac{1}{\sqrt{2}}(|\downarrow\rangle|D\rangle + e^{i\phi}|\uparrow\rangle|U\rangle)
\end{equation}
where $\phi\equiv\delta_Ht_H-\delta_Vt_V-(\delta_H-\delta_V) T+\phi_0$ is the relative phase in the heralded ion-atom entanglement. Here $\delta_H$ and $\delta_V$ denote the frequency differences of the horizontal and vertical components for photons from both sides, $t_H$ and $t_V$ denote the detector click times for H$_\text{left}$ and V$_\text{left}$, $T$ denotes the beginning of the ion state readout, and $\phi_0$ denotes a constant phase independent of the photon arrival time.

Since we have carefully matched the frequency of the vertical components of the photons from both sides, we achieve $\delta_V=0$ and $\delta_H=11.65\,$MHz in our experiment. Subsequently, we fix the time interval between the ion state readout and the click time of detector H$_\text{left}$ to a constant value $T_c$. Through these efforts, the phase $\phi=-\delta_HT_c+\phi_0$ becomes a constant value independent of the random photon arrival times $t_H$ and $t_V$. We then set the phase $\phi$ to $0$ by adjusting the delay time $T_c$, and the resulting ion-atom state can be expressed as:
\begin{equation}
|\Psi_{\text{IA}}\rangle=\frac{1}{\sqrt{2}}(|\downarrow\rangle|D\rangle + |\uparrow\rangle|U\rangle)
\end{equation}

Finally, we characterize the fidelity of the heralded ion-atom entangled state. For the trapped-ion node, we employ $729\,$nm quadrupole transitions and fluorescence detection to perform state readout~\cite{multiplexed}. For the atomic node, we retrieve the two atomic memory modes into an idler photon using a read pulse and measure the state via a combination of a pair of waveplates and a polarizer~\cite{lattice}. Prior to the ion-atom entanglement, we first measure the fidelities of the ion-photon and atom-photon entangled states, which are $96.5\pm0.4\%$ and $84.6\pm0.7\%$, respectively, as illustrated in Fig.~4a and 4b. We characterize the heralded ion-atom entanglement fidelity with respect to the Bell state $|\Psi^+\rangle$ by measuring three correlations $XX$, $YY$, and $ZZ$, as shown in Fig.~4c. The measured correlation functions yield an entanglement fidelity of $\mathcal{F}=\frac{1}{4}(1+XX+YY-ZZ)=75.0\pm4.1\%$, with an ion-atom entangling rate of $0.03\,$s$^{-1}$. It is worth noting that a major source of infidelity is the noise during the readout of the atomic ensemble quantum memory into an idler photon. If readout imperfections are corrected for~\cite{baonature2020}, we estimate the fidelity of the heralded ion-atom entanglement can be improved to about $85\%$.  Furthermore, we find that the infidelity of the ion-atom entanglement is largely dominated by the infidelities of the ion-photon and atom-photon entangled states, with minor contributions from other processes. This demonstrates the effectiveness of the dual-color Bell state measurement scheme we developed. Thanks to the constant phase $\phi=0$ enabled by the dual-color Bell state measurement, we can utilize all the coincidence events within a full $40\,$ns acceptance window to herald the ion-atom entanglement, which improves the heralding efficiency significantly. The measured fidelity as a function of the maximum arrival time difference is illustrated in Fig.~4d. The nearly constant fidelity under different ranges of $|t_H-t_V|$ verifies that the heralded ion-atom entanglement is largely uninfluenced by the random photon arrival times in the heralding process.

\subsection*{Conclusion and outlook}

In this work, we experimentally realize heralded entanglement between a trapped-ion node and an atomic-ensemble node---the foundational step for building an ion-atom hybrid quantum network. By developing technologies such as photon wavepacket shaping and dual-color Bell state measurement, we successfully interconnect two powerful physical platforms for quantum information processing---trapped-ion and neutral-atom systems. Future work will focus on improving the fidelity of the atom-photon entanglement and implementing ensemble-based gates via Rydberg blockade~\cite{blockade, lilin1, lilin2,xuwenchao}, as well as on developing passively stable interferometers for entanglement generation~\cite{kimble_calcite} and wavelength conversion~\cite{hybrid_multiplexing}. In addition, we can improve the entanglement generation rate via single-photon heralding~\cite{pan_atom,pan_ion}, multiplexing~\cite{hybrid_multiplexing, fast_delivery, 500Hz}, and cavity enhancement~\cite{rempe,230m,rempe_teleportation}. With the technologies developed in this work, we can further replace the atomic ensemble with a tweezer-based single-atom array~\cite{fiber_array}, which has recently emerged as a powerful physical platform for various tasks. Moreover, other systems such as color centers and solid-state spin ensembles can also be interconnected via the technologies developed here. Therefore, we believe that the demonstration of heralded ion-atom entanglement and the technologies developed in this work will make a significant contribution to the realization of the future hybrid quantum network and global quantum internet.\\

\textbf{Note added:} During the preparation of this manuscript, we became aware of a related work by IonQ Inc.~\cite{IonQ}, which entangles a trapped ion and a silicon-vacancy center via a different scheme.

\noindent\textbf{Data availability} The data that support the findings of this study are available from the
corresponding authors upon request.

\noindent\textbf{Acknowledgements}
This work is supported by Innovation Program for Quantum Science and
Technology (No.2021ZD0301604, No.2021ZD0301102), the Tsinghua University
Initiative Scientific Research Program and the Ministry of
Education of China through its fund to the IIIS. Y.F.P. acknowledges
support from the Dushi Program from Tsinghua University.

\end{document}